\documentclass[11pt,reqno]{amsart}
\usepackage[pdfstartview=FitB]{hyperref}
\usepackage{natbib}
\usepackage{geometry}                % See geometry.pdf to learn the layout options. There are lots.
\usepackage{amsmath,amssymb,amsthm}
\theoremstyle{plain}

\theoremstyle{definition}

\newcommand{\citeasnoun}{\cite}

\begin{document}

\title[Post-Selection and Post-Regularization]{Post-Selection and Post-Regularization Inference in Linear Models with Many Controls and Instruments}
\author{Victor Chernozhukov, Christian Hansen, and Martin Spindler}\thanks{Chernozhukov: Massachussets Institute of Technology, 50 Memorial Drive, E52-361B, Cambridge, MA 02142, vchern@mit.edu.
Hansen: University of Chicago Booth School of Business, 5807 S. Woodlawn Ave., Chicago, IL 60637, chansen1@chicagobooth.edu. Spindler:  Munich Center for the Economics of Aging, Amalienstr. 33, 80799 Munich, Germany, spindler@mea.mpisoc.mpg.de. }
%\pubMonth{Month}
\date{January 5, 2015}
%\pubYear{Year}
%\pubVolume{Vol}
%\pubIssue{Issue}
\maketitle

\begin{quote}
Abstract. In this note, we offer an approach to estimating structural parameters in the presence of many instruments and controls based on methods for estimating sparse high-dimensional models.  We use these high-dimensional methods to select both which instruments and which control variables to use.  The approach we take extends \citeasnoun{BCCH12}, which covers selection of instruments for IV models with a small number of controls, and extends \citeasnoun{BelloniChernozhukovHansen2011}, which covers selection of controls in models where the variable of interest is exogenous conditional on observables, to accommodate both a large number of controls and a large number of instruments.  We illustrate the approach with a simulation and an empirical example.  Technical supporting material is available in a supplementary appendix. \\

\bigskip

\noindent Publication: American Economic Review 2015, Papers and Proceedings. \\

\noindent Online Appendix:  Post-Selection and Post-Regularization Inference: An Elementary, General Approach.\\

\end{quote}

\section{Model and Estimation Approach}

Consider the linear IV model
\begin{align}
\label{ystructure}
y_i &= \alpha_0 d_i + x_i' \beta_0 + \varepsilon_i, \\
\label{dzx}
d_i &= x_{i}'\gamma_0 + z_{i}'\delta_0 + u_i,   
\end{align}
with $\textrm{E}[(z_i',x_i')'\varepsilon_i] = \textrm{E}[(z_i',x_i')'u_i] = 0$.  $d_i$ is the scalar endogenous variable and $\alpha$ the coefficient of interest, $x_i$ is a $p_n^x$-vector of exogenous control variables, $z_i$ is a $p_n^z$-vector of instruments, $n$ is the sample size, and $p_n^x \gg n$ and $p_n^z \gg n$ are allowed.  Extension to the case where $d_i$ is a vector is straightforward and omitted for simplicity.  We may have that $z_i$ and $x_i$ are correlated so that $z_i$ are only valid instruments after controlling for $x_i$; specifically, we let $z_i = \Pi x_i + \zeta_i,$ for $\Pi$ a $p_n^z \times p_n^x$ matrix and $\zeta_i$ a $p_n^z$-vector of unobservables with $\textrm{E}[x_i \zeta_{i}'] = 0$.  Substituting this expression for $z_i$ as a function of $x_i$ into (\ref{dzx}) and then further substituting into (\ref{ystructure}) gives a system for $y_i$ and $d_i$ that depends only on $x_i$:
\begin{align}
\label{yx}
y_i &= x_{i}'\theta_0 + \rho^y_i, \\
\label{dx}
d_i &= x_{i}'\vartheta_0 + \rho^d_i,
\end{align}
with $\textrm{E}[x_i \rho^y_i] = 0$ and $\textrm{E}[x_i \rho^d_i] = 0$.  This model includes the many instruments and small number of controls case by setting $p_n^x \ll n$ and can accommodate the exogenous case by setting $p_n^z = 0$ and imposing the additional condition $\textrm{E}[d_i \varepsilon_i] = 0$.  
%The main parameter here is $\alpha$ and the nuisance parameter is $$\eta = ({\theta}', {\vartheta}', \gamma', \delta')'.$$

Because the dimension of $\eta_0 = ({\theta}'_0, {\vartheta}_0', \gamma_0', \delta_0')'$ may be larger than $n$, informative estimation and inference about $\alpha_0$ is impossible without imposing restrictions on $\eta_0$.  For simplicity, we provide discussion under the assumption of exact sparsity and present a generalization to approximate sparsity in the supplemental material.  Specifically, we assume that $$\|\eta_0\|_0  \leq{s_n}, \quad  s_n^2 \log(p_n^z + p_n^x)^3/n \rightarrow 0,$$
where $\|\eta_0\|_0$ denotes the number of non-zero elements of $\eta_0$.  That is, sparsity requires that, among the $p_n^x + p_n^z$ observed variables, the number of variables with non-zero coefficients is small relative to the sample size.  This assumption then reduces the problem of estimating $\alpha$ to a problem of finding which instruments and controls to use in equations (\ref{ystructure}) and (\ref{dzx}).  

The problem that arises is that variable selection techniques are not perfect and are prone to making selection mistakes.  There are two kinds of selection mistakes: A variable may be deemed relevant when in fact it has a zero coefficient and thus has no true explanatory power, or a variable may be dropped from the model despite having a non-zero coefficient.  Both types of mistakes may detrimentally affect post-model-selection estimators and inference for $\alpha$.  When irrelevant variables are spuriously included after being deemed predictive from looking at the data, overfitting occurs and importantly the spuriously included variables are those most correlated to the noise in the sample due to data-snooping which introduces a type of ``endogeneity'' bias.  When relevant $x$ variables are excluded, one is left with standard omitted variables bias.  When relevant $z$ variables are excluded, one loses identification power.  This last concern can be dealt with through appropriate use of weak identification robust inference as in \citeasnoun{BCCH12}.

The first type of mistake, the spurious inclusion of irrelevant variables, can be avoided through the use of modern, principled data-mining methods.  For example, we use the Lasso with tuning parameters chosen as in \citeasnoun{BCCH12}, and many other options are available.  These methods differ from the unprincipled data-snooping that many economists associate with the term data-mining.  Specifically, modern data-mining denotes a principled search for "true" predictive power that guards against false discovery and overfitting, does not erroneously equate in-sample fit to out-of-sample predictive ability, and accurately accounts for using the same data to examine many different hypotheses or models.

Of course, guarding against the first type of error comes at the cost of needing to acknowledge that the exclusion of relevant variables is likely to occur.  While sensible approaches such as Lasso will accurately find strong predictors, one can show that such procedures have non-negligible probability of missing predictors with small but non-zero coefficients.  Exclusion of such predictors can have substantive impacts on inference for parameters of interest such as $\alpha$ in our model; see, for example, \citeasnoun{leeb:potscher:pms}.  To overcome this difficulty, one needs to base estimation and inference on procedures that are robust to this type of model selection mistake.  One such approach relies on using estimating equations that are locally insensitive to this type of mistake, termed orthogonal moment functions in \citeasnoun{BCFH:Policy}.

In the IV model with many instruments and controls, such a moment condition is given by
\begin{align}\label{moment}
M(\alpha_0; \eta_0) = 0,  \ M(\alpha, \eta) := \textrm{E} \left[ \psi_i(\alpha, \eta)\right]
\end{align}
where $\psi_i(\alpha, \eta) = (\tilde \rho^y_i- \tilde \rho^d_i  \tilde \alpha)\tilde v_i $ for $\eta := (  \theta',  \vartheta', \gamma',  \delta')'$, $\tilde \rho^y_i:= y_i - x_{i}'\theta$, $\tilde \rho^d_i:= d_i-x_{i}' \vartheta$, and $\tilde v_i := x_{i}'  \gamma+ z_{i}' \delta - x_{i}'\vartheta$.  When we set $\tilde \eta = \eta_0$, we have $\tilde \rho^y_i = \rho^y_i = y_i - x_{i}'\theta_0$, $\tilde \rho^d_i= \rho^d_i= d_i-x_{i}'\vartheta_0$, and $\tilde v_i = v_i :=x_{i} '\gamma_0+ z_{i}'\delta_0 - x_{i}'\vartheta_0= \zeta_i'\delta_0$.  

We can see that small selection errors will have relatively little impact on estimation of $\alpha_0$ by noting that the following orthogonality condition holds:
\begin{align}
\label{orthogonality}\frac{\partial }{\partial \eta }  M(\alpha_0, \eta) \Big \vert_{\eta= \eta_0} =0.
\end{align}  
In other words, missing the true value $\eta_0$ by a small amount does not invalidate the moment condition.  Thus, estimators
$\hat \alpha$ of $\alpha_0$ based on the empirical analog of (\ref{moment}),
\begin{align}
\label{emp moment}
\hat M(\hat \alpha, \hat \eta) = 0
\end{align}
with $\hat M(\alpha, \eta)  := n^{-1}\sum_{i=1}^n  \left[ \psi_i(\alpha, \eta)\right],$ 
can be shown to be ``immunized'' against small selection mistakes. 
%(Here and below $\En$ abbreviates the average $n^{-1}\sum_{i=1}^n$ over index $i$.)  
%The first use of orthogonal moment conditions in obtaining valid inference on low-dimensional components in sparse high-dimensional models is given in \citeasnoun{BellChernHans:Gauss} in the context of Gaussian IV regression with a small number of controls. See also \citeasnoun{BCFH:Policy} which contains the current most general formulation along with a number of estimation and inference results.
See \citeasnoun{BCFH:Policy} for a general formulation of orthogonal moment funtions for use in sparse high-dimenionsal models and a number of estimation and inference results.

Note that operationally using the empirical version of (\ref{moment}) to estimate $\alpha_0$ is equivalent to using the usual IV regression of $\rho^y$ on $\rho^d$ using $v$ as instruments. Based on this argument, we suggest the following algorithm for estimating $\alpha_0$ based on the ``double-selection'' strategy of \citeasnoun{BelloniChernozhukovHansen2011}. \\

\textrm{Algorithm 1.} \textit{ (1) Do Lasso or Post-Lasso Regression of $d_i$ on $x_i, z_i$ to obtain $\hat \gamma$ and $\hat \delta$. (2) Do Lasso or Post-Lasso Regression of $y_i$ on $x_i$ to get $\hat \theta$. (3) Do Lasso or Post-Lasso Regression of $\hat d_i = x_i'\hat\gamma + z_i'\hat\delta$ on $x_i$ to get
	 $\hat \vartheta$. (4) Let $\hat{\rho}^y_i := y_i - x_i'\hat \theta$, $\hat{\rho}^d_i := d_i - x_i'\hat \vartheta$, and 
	$\hat{v}_i := x_i'\hat \gamma + z_i'\hat \delta - x_i'\hat \vartheta$. Get estimator $\hat \alpha$ from (\ref{emp moment}) by using standard IV regression of $\hat{\rho}^y_i$ on $\hat{\rho}^d_i$ with $\hat{v}_i$ as the instrument.  Perform inference on $\alpha_0$ using $\hat \alpha$ 
	or the associated score statistic and conventional heteroscedasticity robust standard errors.} \\

The following result summarizes the properties of $\hat\alpha$ obtained from Algorithm 1. \\

\textrm{Proposition 1.} \textit{Under the stated sparsity and other regularity conditions, the estimator $\hat\alpha$ defined in Algorithm 1 satisfies $\sqrt{n} (\hat \alpha - \alpha_0) \rightsquigarrow \mathcal{N}(0, V)$
where  $V = \textrm{E}[v_i^2]^{-2}  \textrm{E}[ \psi_i(\alpha_0, \eta_0)^2]$. The score statistic
$C(\alpha_0) = n|\hat M(\alpha_0, \hat \eta)|^2 / (n^{-1}\sum_{i=1}^n \psi^2_i(\alpha_0, \hat \eta)) $ satisfies
$C(\alpha_0) \rightsquigarrow \chi^2(1)$.   Confidence intervals based on these two results are uniformly valid for inference about $\alpha_0$ over a large class of models.} \\

The supplementary material provides a precise statement and proof. The theoretical results do not depend on whether the Lasso estimator or the Post-Lasso estimator of \citeasnoun{BC-PostLASSO} is used.  In the results reported in this paper, we use the Post-Lasso estimator.  Note that there are other algorithms that would yield similar asymptotic properties.  For example, one could follow the double-selection strategy more closely by running Lasso regression of $d_i$ on $x_i$ and $z_i$, Lasso regression of $d_i$ on $x_i$, Lasso regression of $y_i$ on $x_i$, and then forming a 2SLS estimator using instruments selected in the first step and controlling for the union of controls selected in the three Lasso steps.

\section{Simulation Example}

To illustrate the preceding discussion, we report results from a small simulation experiment.  Data were generated from the model given in Section 2 with $n = 200$, $p_n^x = 300$, and $p_n^z = 150$.  Other parameter values were chosen so that the infeasible, optimal instruments are ``strong'', perfect model selection is impossible, and the sparse model provides a good approximation.  Further details are available in the supplementary material.  

We provide results for four different estimators - an infeasible Oracle estimator that knows the nuisance parameters $\eta$ (Oracle), two naive estimators, and the ``Double-Selection'' estimator.  The first naive estimator follows Algorithm 1 but replaces Lasso/Post-Lasso with stepwise regression with p-value for entry of .05 and p-value for removal of .10 (Naive 1).  It is well-known that this procedure fails to control model selection mistakes in which irrelevant variables are included.  The second naive estimator estimates the high-dimensional nuisance functions using Post-Lasso but uses the moment condition $\textrm{E} \left[ (\rho_i^y -\rho_i^d\alpha) (x_i'\delta + z_i'\gamma) \right] =0$ (Naive 2).  This moment condition does not satisfy the orthogonality condition described above, though estimation and inference about $\alpha_0$ using this condition will be valid when perfect model selection for the regression of $y$ on $x$ and $d$  on $x$  is possible.

%\begin{center}
%\includegraphics[width=\columnwidth]{SimulationFigure}
%\captionof{figure}{Sampling distributions of standardized estimators from simulation experiment.}
%\end{center}
%
%\begin{figure}
%\includegraphics[width=\columnwidth]{SimulationFigure}
%\caption{Sampling distributions of standardized estimators from simulation experiment.}
%\end{figure}

%We display the simulation results in Figure 1.  For each estimator, we plot the simulation estimate of the sampling distribution of the estimator centered around the true parameters and scaled by the estimated standard error.  With this standardization, usual asymptotic approximations would suggest that these curves should line up with a N(0,1) which is displayed as the bold solid line in the figure.  We can see that the Oracle estimator and the Double-Selection estimator are centered correctly with the Oracle lining up well with N(0,1) and the Double-Selection estimator exhibiting a small amount of excess dispersion.  The first naive estimator which fails to control spurious inclusion of variables is centered far from zero and more widely dispersed than the other estimators.  The second naive estimator has center closer to zero than the first but which is still noticeably shifted out relative to the principled approach.  

We report the median bias (Bias), median absolute deviation (MAD), and size of 5\% level tests (Size) obtained from 1000 simulation replications for each procedure.  For the Oracle, we have Bias of .006, MAD of .095, and Size of .043.  For Naive 1, Bias, MAD, and Size are .160, .227, and .302 respectively; and Bias, MAD, and Size are respectively .035, .103, and .095 for Naive 2.  Finally, the Double-Selection approach gives Bias of .021, MAD of .099, and Size of .054.  

These results correspond to the discussion in Section I.  The first naive, unprincipled procedure fails to control spurious inclusion of irrelevant variables and performs quite poorly relative to the other three approaches.  The second naive procedure can be shown to be formally valid when perfect model selection is possible and performs relatively well in terms of MAD.  However, the asymptotic approximation under perfect model selection provides a misleading approximation to the true sampling distribution as evidenced by the size distortion.  %This distortion is the result of basing the estimator and inference on a moment condition that is not first-order insensitive to small perturbations in the nuisance parameters and the well-known fact that principled model selection procedures will fail to include variables with moderately sized coefficients with high-probability.  
Finally, we see that basing estimation and inference on a principled variable selection procedure and moment conditions that are immunized against small model selection mistakes produces an estimator that performs well relative to the infeasible Oracle in terms of both estimation and inference performance as measured by MAD and Size.

\section{Empirical Example}

We conclude with a brief empirical example where we estimate the coefficients in a simple model of demand for automobiles.  We use the data and basic strategy of \citeasnoun{BLP}.  For simplicity, we consider the most basic specification 
\begin{align*}
\log(s_{it}) - \log(s_{0t}) &= \alpha_0 p_{it} + x_{it}'\beta_0 + \varepsilon_{it} \\
p_{it} &= z_{it}'\delta_0 + x_{it}'\gamma_0 + u_{it}
\end{align*}
where $s_{it}$ is the market share of product $i$ in market $t$ with product 0 denoting the outside option, $p_{it}$ is price and treated as endogenous, $x_{it}$ are observed included product characteristics, and $z_{it}$ are instruments.  One could also consider allowing random coefficients and adapting the variable selection procedures to this setting; see \citeasnoun{BLPLASSO}.  

In their basic results, \citeasnoun{BLP} use five variables in $x_{it}$: a constant, an air conditioning dummy, horsepower divided by weight, miles per dollar, and vehicle size.  
%They argue that characteristics of other products provide valid instruments for price which leaves a very high-dimensional set of instruments.  
%Based on intuition and an exchangeability argument, \citeasnoun{BLP} choose to use each element of $x$ summed across all other products produced by the same firm as produces product $i$ and each element of $x$ summed across all products produced by firms excluding the firm that produces product $i$ as instruments for $p_{it}$.  This choice yields a set of 10 instruments.  
They argue that characteristics of other products provide valid instruments for price and choose 10 instruments for $p_{it}$ based on intuition and an exchangeability argument.  The first five instruments are formed by deleting product $i$ and then summing each characteristic in $x$ across all remaining products produced by product $i$'s firm.  The other five instruments are similarly constructed by deleting all products from product $i$'s firm and then summing each characteristic in $x$ across all remaining products.
Using these controls and instruments, the 2SLS estimate of $\alpha$ is -.142 with an estimated standard error of .012.  One might compare this to the OLS estimate obtained treating price as exogenous given the five controls listed above which is -.089 with estimated standard of .004.

It is interesting to note that \citeasnoun{BLP} state, ``The choice of which attributes to include in the utility function is, of course,
ad hoc'' (p. 872).  They similarly note that one could have considered additional instruments such as higher order terms \cite[p. 861]{BLP}.  The high-dimensional methods outlined in this paper offer one strategy to help address these concerns which complements the well-founded economic intuition motivating the authors' choices.  We apply our outlined methods in two scenarios.  In the first, we apply the method using just the original five controls and 10 instruments.  In the second, we augment the set of potential controls with a time trend, quadratics, and cubics in all continuous variables, and all first order interactions and then use sums of these characteristics as potential instruments following the original strategy.  These additions give a total of 24 $x$-variables and 48 potential instruments.  We include the intercept in all models and select over the remaining variables.

In both cases, the results suggest demand is more elastic than indicated in the baseline results.  After selection using only the original variables, we estimate the price coefficient to be -.185 with an estimated standard error of .014.  In this case, all five controls are selected in the log-share on controls regression, all five controls but only four instruments are selected in the price on controls and instruments regression, and four of the controls are selected for the price on controls relationship.  The difference between the baseline results is thus largely driven by the difference in instrument sets.  The change in the estimated coefficient is consistent with the wisdom from the many-instrument literature that inclusion of irrelevant instruments biases 2SLS toward OLS.  

With the larger set of variables, our post-model-selection estimator of the price coefficient is -.221 with an estimated standard error .015.  Here, we see some evidence that the original set of controls may have been overly parsimonious.  In the log-share on controls regression, we have that eight control variables are selected; and we have seven controls and only four instruments selected in the price on controls and instrument regression.  We also have that 13 variables are selected for the price on controls relationship.  The selection of these additional variables suggests that there is important nonlinearity missed by the baseline set of variables.

Finally, we note that in terms of own-price elasticities, the results become more plausible as we move from the baseline results to the results based on variable selection with a large number of controls.  Recall that facing inelastic demand is inconsistent with profit maximizing price choice within the present context, so theory would predict that demand should be elastic for all products.  However, the baseline point estimates imply inelastic demand for 670 products.  Using the variable selection results provides results closer to the theoretical prediction.  The point estimates based on selection from only the baseline variables imply inelastic demand for 139 products, and we estimate inelastic demand for only 12 products using the results based on selection from the larger set of variables.  Thus,
the new methods provide the most reasonable estimates of own-price elasticities. Of course, the simple specification above suffers from the usual drawbacks of the logit demand model, but the example illustrates how the application of the methods outlined in this note may be used in estimation of structural parameters in economics and add to the plausibility of the resulting estimates.

\section{Conclusion}

A great deal of empirical economic research aiming to estimate causal or structural effects depends on using the right set of controls and instruments.  The need for formal methods that perform this model selection and inference procedures that remain valid following model selection is likely to increase in importance as data sets become richer.  We have outlined one simple approach that can be used in an instrumental variables model with many instruments and controls that extends \citeasnoun{BCCH12} and \citeasnoun{BelloniChernozhukovHansen2011}.  The approach relies on an approximate sparsity assumption and the use of high-quality variable selection procedures coupled with the use of appropriate moment functions.  These ideas follow from the general framework considered in \citeasnoun{BCFH:Policy}.  For more applications of similar ideas in economics, see also \citeasnoun{BaiNg2009a}, \citeasnoun{BellChernHans:Gauss}; \citeasnoun{GautierTsybakovHDIV}; \citeasnoun{BCH2011:InferenceGauss}; and \citeasnoun{BCHK:FE} and references therein.  
%There remain many interesting areas to explore in the use of high-dimensional methods on economics including other types of dimension reduction strategies, applications in other economic models, and the feasibility of valid inference when sparsity assumptions do not hold.

\bibliographystyle{aea}
\bibliography{mybib}

\end{document}